\documentclass[journal,draftclsnofoot,onecolumn]{IEEEtran}

\usepackage{graphicx} % Required for inserting images

\usepackage{amsmath,amssymb,amsthm}
\usepackage{hyperref}
\usepackage{tikz}
\usetikzlibrary{shadows, calc}
\usepackage{pgfplots}
\pgfplotsset{compat=1.18}
\usepackage{pgfplotstable}
\usepgfplotslibrary{colormaps}

\title{Teaching at Scale: Leveraging AI to Evaluate and Elevate Engineering Education}
\author{
J.-F. Chamberland, M. Carlisle, A. Jayaraman, K. R. Narayanan, S. Palsole, K. Watson\\
\IEEEauthorblockA{College of Engineering, Texas A\&M University}}

\begin{document}

\maketitle

\begin{abstract}
Evaluating teaching effectiveness at scale remains a persistent challenge for large universities, particularly within engineering programs that enroll tens of thousands of students.
Traditional methods, such as manual review of student evaluations, are often impractical, leading to overlooked insights and inconsistent data use.
This article presents a scalable, AI-supported framework for synthesizing qualitative student feedback using large language models. The system employs hierarchical summarization, anonymization, and exception handling to extract actionable themes from open-ended comments while upholding ethical safeguards.
Visual analytics contextualize numeric scores through percentile-based comparisons, historical trends, and instructional load. The approach supports meaningful evaluation and aligns with best practices in qualitative analysis and educational assessment, incorporating student, peer, and self-reflective inputs without automating personnel decisions.
We report on its successful deployment across a large college of engineering.
Preliminary validation through comparisons with human reviewers, faculty feedback, and longitudinal analysis suggests that LLM-generated summaries can reliably support formative evaluation and professional development.
This work demonstrates how AI systems, when designed with transparency and shared governance, can promote teaching excellence and continuous improvement at scale within academic institutions.
\end{abstract}

\section{Introduction}

The College of Engineering at Texas A{\&}M University is committed to \emph{Educational Excellence at Scale}, which means ensuring high-quality learning experiences for all students, at all levels, in courses of all sizes (large and small).
As a land-grant institution, Texas A{\&}M prioritizes broad accessibility in service of the public good.
The College currently enrolls approximately 25,000 students in engineering, computing, and engineering technology programs. 
Despite its size, the College maintains a strong focus on instructional quality, with the aspirational goal of cultivating scholarly teaching among all instructors.
To enrich the student learning experience, the College embraces best practices in engineering education, including student evaluations, peer observations, teaching portfolios, and self-reflection~\cite{smith2005pedagogies,ambrose2010learning,felder2024teaching}.
This article examines the role of student evaluation of teaching in large-scale institutions.
At the outset, we emphasize that teaching quality should never be judged solely on student evaluations; rather, this article explores an approach to enhance their value within large-scale institutions.

End-of-term student evaluations generate vast amounts of data, particularly in large engineering programs, where tens of thousands of students provide numeric ratings and open-ended comments each semester.
While instructors are expected to review their own student comments, the volume of qualitative feedback makes it challenging for peer evaluators and department heads to read every comment across their departments, and essentially impractical for college-level administrators to do so for every instructor in their unit.
At our institution, end-of-term evaluations can generate hundreds of thousands of data points each semester within engineering.
As a result, important insights from written comments are often overlooked, raising concerns about how to use this feedback fairly and effectively to support faculty development.

To address this challenge, the College has implemented an AI-driven system to summarize student comments.
This workflow builds on earlier pilot studies~\cite{katz2024using}, including our own proof-of-concept, and it leverages large language models (LLMs).
The system generates thematic summaries based on six key institutional questions related to instructional quality.
It incorporates hierarchical text summarization, identity anonymization, and exception tagging for inappropriate content (e.g., flagging harassing remarks).
The summaries are presented alongside visual analytics, including historical trends, teaching loads, and percentile rankings, to provide context. 
Based on the evidence, faculty can be categorized as ``award-worthy,'' ``meets expectation,'' or ``needs improvement'' enabling leadership to target reviews and allocate resources strategically.
The goal is to foster a supportive, improvement-focused culture rather than a punitive evaluation process~\cite{hattie2007power}.

The central question we aim to address is: \textbf{How can AI-driven summarization and evaluation systems enhance instructional quality and support faculty development at scale in engineering-focused institutions?}
In this article, we explore this question by synthesizing relevant literature and established best practices.
We begin by reviewing scholarship on scalable, data-informed teaching evaluations.
Next, we examine how AI summarization techniques align with qualitative analysis standards and ethical considerations~\cite{davison2024ethics}.
We then discuss how to ensure context and fairness in interpreting student feedback, including factors such as instructor identity, class size, and course structure.
We consider the role of visual analytics, real-time feedback, and percentile-based scoring in shaping instructor motivation and guiding departmental decision-making.
We outline strategies for validating and benchmarking the AI-supported system against traditional evaluation models, including peer and supervisor reviews.
Finally, we address the risks and limitations of incorporating AI into teaching evaluation and propose ways to mitigate these through thoughtful design, transparency, and governance, before turning to broader implications for institutional policy and professional development.

\section{Literature Review -- Scalable and Data-Informed Teaching Evaluations}

Student evaluations of teaching (SET) are now a standard feature of higher education, commonly used in quality assurance processes and to inform decisions about faculty promotion, teaching awards, and instructional development~\cite{murray1984impact,marsh1987students,murray1997does,zabaleta2007use}.
According to the Texas Administrative Code (Title~19, Part~1, Chapter~4, Subchapter~N, Rule~4.228), Texas state law requires public institutions of higher education, other than a medical and dental unit, to conduct end-of-course student evaluations of faculty for each undergraduate classroom course and to make these evaluations
publicly available on the institution's website.
In contrast, some literature questions the validity, potential biases, and overall utility of these evaluations.
Although student surveys can offer valuable insight into the learning experience, researchers have noted a weak correlation between SET scores and actual learning outcomes \cite{clayson2009student,deslauriers2019measuring}.
In other words, high student satisfaction does not necessarily indicate high levels of student learning, raising important concerns about the use of SET scores as direct indicators of teaching effectiveness.
Moreover, SET results are often shaped by factors beyond the instructor's control.
Course and contextual variables, such as class size, course level, and whether a course is required or elective, can significantly influence student ratings~\cite{feldman1984class,bedggood2012university}.
For example, large lecture courses and required core classes tend to receive lower evaluations than small elective ``special topics'' courses~\cite{dee2007student}.
These patterns underscore the importance of contextualizing raw evaluation metrics to ensure a fair and accurate interpretation.

In response to these concerns, scholars and institutions have proposed more holistic, data-informed evaluation models.
Best practices call for using multiple measures of teaching effectiveness instead of over-relying on student surveys alone.
For instance, Brookfield's critical reflection model advocates that teachers examine their practice through ``four lenses'': the students' eyes (student feedback), colleagues' perceptions (peer review), personal experience (self-reflection), and theory/research literature~\cite{brookfield2017becoming}.
This approach recognizes that each source of evidence provides a different perspective, and only by considering them together can one form a balanced view of teaching quality.
In line with this, many universities have introduced peer observations, teaching portfolios, and self-reflection statements to complement student input.
For example, faculty might write a reflective statement explaining how they have responded to recurring student comments, and peers may conduct classroom observations to provide constructive critique.
The College of Engineering guidelines explicitly state that ``In the comprehensive evaluation of faculty teaching performance, it is crucial to look beyond student feedback and incorporate a multifaceted assessment approach.'' This multifaceted approach ensures a balanced and in-depth evaluation of teaching effectiveness, recognizing the importance of continuous improvement and innovation in educational practices.

Another thread in the scholarship of teaching and learning (SoTL) literature focuses on making sense of large-scale evaluation data for actionable insights.
Traditional qualitative analysis of open-ended comments (e.g., manual coding of themes) is rigorous but labor-intensive and hard to scale beyond small samples.
Recent studies in engineering and higher education have explored the application of text mining and machine learning to student feedback~\cite{sun2023using,katz2024using}.
For example, topic modeling techniques like latent Dirichlet allocation (LDA) have been used to extract common themes from thousands of course comments~\cite{gottipati2018latent}.
One large study analyzed student comments from hundreds of courses using LDA and found that integrating qualitative topics with quantitative scores significantly enhanced the evaluation of teaching performance.
The open-ended comments yielded concrete, specific feedback often beyond the scope of the numeric survey dimensions.
In fact, the extracted topics were ``more specific than the quantitative dimensions of the survey,'' providing very concrete feedback for professors and informing targeted training programs.
Crucially, about half of the topics identified were ``actionable'' items not tied to a professor's fixed traits (e.g.
course pace or materials), which means they highlighted areas an instructor could realistically improve.
These granular insights, e.g., recurring mentions of unclear assignments or insufficient office hours, can directly spur teaching development initiatives.
Student comments prompted evidence-based reflections by faculty committees that led to concrete actions to improve course quality.
In their case, inviting students to freely comment gave the program an ``extensive view'' of teaching in the degree, and the feedback was used formatively for changes to instruction and curriculum.
These findings reinforce that open-ended student feedback, when properly analyzed, is a powerful tool for continuous improvement, one that can complement the ``blunter'' numerical ratings.

Finally, the teaching excellence frameworks in the literature emphasize reflective improvement and evidence-based practice.
For instance, Kember's model of teaching and learning suggests that measuring teaching quality should account for instructors' conceptions of teaching and their reflection on feedback, not just student satisfaction metrics~\cite{kember2012evaluating}.
Similarly, models like Marsh's Teacher Effectiveness framework and Kember's teaching excellence dimensions have long argued for using multiple indicators and providing faculty with rich feedback to refine their teaching strategies~\cite{marsh1997making,spooren2013validity}.
In essence, the scholarship calls for data-informed yet context-sensitive approaches: using the wealth of data (including comments, survey scores, peer input, etc.) to guide improvement, but doing so in a way that is fair, reliable, and formative rather than merely judgmental.
This sets the stage for AI-supported systems, when designed well, to aggregate and analyze large-scale feedback in service of those goals~\cite{mikolov2013distributed,sutskever2014sequence,vaswani2017attention,devlin2019bert,NEURIPS2020_1457c0d6}.

\section{AI Summarization versus Qualitative Analysis}

Adopting AI to summarize student comments raises an important question: \textbf{Can AI techniques uphold the best practices of qualitative analysis and meet the ethical standards of teaching evaluation?}
Traditional qualitative methods, such as thematic coding and content analysis, involve systematically reading comments, identifying themes, and distilling insights while preserving the authentic voice of respondents.
Analysts often ensure trustworthiness through practices like inter-rater reliability (multiple reviewers agreeing on themes) and by situating quotes within appropriate context.
Ethical standards further demand confidentiality (protecting identities of students and instructors) and neutrality (avoiding bias in interpreting feedback).

When implemented carefully, AI-driven summarization, particularly with large language models (LLMs), can align with these principles~\cite{johri2023generative,morgan2023exploring,turobov2024using,wachinger2025prompts,naeem2025thematic}.
Our system uses a hierarchical summarization approach that parallels human coding: the code first groups comments by topic, summarizes each cluster, and then synthesizes higher-level summaries.
This mirrors the process of identifying sub-themes and overarching themes in manual analysis.
By preserving structure, the LLM generates summaries that reflect distinct dimensions of teaching (e.g., lecture clarity, course organization, critical thinking) rather than collapsing everything into a single blended paragraph.
This helps retain nuance and allows for more actionable insights.
In essence, the AI compresses large volumes of data while preserving key categories and diverse perspectives, much like a human qualitative researcher would.

We also incorporate comment anonymization and exception tagging as preprocessing steps, consistent with ethical best practices~\cite{mehrabi2021survey}.
Anonymization removes or masks any personally identifying information (e.g., names of instructors or students) from the AI's input.
This protects confidentiality and reduces the risk of model bias, thus preventing the model from forming evaluative judgments based on repeated exposure to a particular name.
Instead, the AI focuses on the substance of the feedback.
Exception tagging flags comments containing hate speech, personal attacks, or harassment for administrative review, rather than including them in the summary or final report.
This mirrors a human analyst's judgment to set aside inappropriate or harmful remarks, ensuring that they do not distort the tone of the summary while still acknowledging their occurrence.
By isolating extreme or malicious comments, the system maintains a constructive focus on pedagogical improvement, in line with the ethical intent of student evaluations.

A key guideline emerging in higher education is that AI-generated summaries should inform human judgment, not replace it.
The College has adopted this principle in its guide on using artificial intelligence to support faculty evaluations.
AI may be used as an initial step to help summarize large volumes of qualitative data (such as student comments), but human reviewers are expected to apply their own insight and discretion in interpreting the results.
The AI's role is deliberately narrow: it generates digests to aid in reviewing feedback, but it does not assign evaluative labels such as ``outstanding'' or ``needs improvement.'' Decisions, such as nominations for teaching awards or recommendations for professional development, should be based on a holistic review of evidence, including peer evaluations, self-reflections, and teaching context.
This division of labor reflects best practices in qualitative analysis: AI contributes to descriptive synthesis, while \textbf{evaluative and inferential judgments remain the responsibility of human reviewers}.
In practice, the system helps surface cases that merit closer attention during the review process, supporting more targeted and informed evaluations without automating personnel decisions.

Another way in which AI summarization aligns with qualitative methods is by potentially improving consistency and reducing individual bias during the initial analysis stage~\cite{kasneci2023chatgpt,johri2023generative,zhang2023redefining,farrokhnia2024swot}.
Human readers are susceptible to confirmation bias or emotional responses~\cite{klayman1995varieties,nickerson1998confirmation}.
For example, a particularly harsh comment may linger in memory and overshadow numerous positive ones. 
In contrast, an AI system, devoid of emotion, can provide a more balanced aggregation of feedback. 
Fuller et al.\ examined the use of ChatGPT to analyze course evaluations and found that AI identified themes that closely aligned with those identified by human coders, and in some cases, added helpful detail.
Notably, the AI was not influenced by the emotional intensity of negative comments in the way instructors sometimes are; it treated all input analytically, which may contribute to more impartial summaries of areas for improvement.
Instructors in the study reported experiencing anxiety and emotional strain when manually reviewing feedback, whereas the AI produced objective summaries that distilled critical points without emotional framing. 
This suggests that AI can replicate aspects of systematic coding while avoiding halo effects or defensiveness, offering a straightforward presentation of critical feedback. 
That said, the study also noted that ChatGPT occasionally generated overly granular sub-themes that human reviewers might deem less relevant, highlighting the need for human oversight to ensure that summaries remain focused and meaningful.
In essence, AI can efficiently perform the heavy lifting of organizing and synthesizing student comments, but human reviewers are essential for validating the relevance and interpretability of the results.

In terms of efficiency, AI summarization significantly outpaces traditional methods, making large-scale evaluation more feasible.
Fuller et al.\ reported that instructors spent an average of 27.5 ($\pm 15$) minutes analyzing a set of course comments~\cite{fuller2024exploring}, whereas ChatGPT completed the same task in seconds with appropriate prompting.
When scaled across hundreds of courses, this time savings is substantial, freeing faculty committees and administrators from hours of manual labor and allowing them to focus on interpreting results and planning interventions.
This is very much aligned with education excellence at scale.
Still, speed should not come at the expense of rigor.
Researchers emphasize that successful use of large language models (LLMs) requires thoughtful prompt design and careful data preprocessing.
For example, chunking text to comply with token limits and crafting prompts that instruct the model to be accurate, concise, and neutral are critical steps.
These tasks (cleaning data, batching inputs, and refining prompts) introduce some overhead, but this is managed with automation in our implementation~\cite{mehta1998computerized}.

Data privacy and security are also essential ethical considerations.
Many AI summarizers rely on public LLMs, raising concerns about uploading sensitive student feedback to external servers.
In our deployment, we mitigate this by using a subscription-based (application programming interface) API set, which preserves privacy, and by anonymization of the data prior to calling the API~\cite{brown2022does}.
 Safeguarding student comments, which may include personal anecdotes, is paramount.
Responsible data stewardship is a cornerstone of ethical teaching evaluation.

AI summarization can align with qualitative best practices when it follows structured, hierarchical methods that preserve the granularity of the data~\cite{zhang2023redefining,drinkwater2025expanding}.
It should also incorporate anonymization and input filtering to protect confidentiality and maintain a focus on constructive feedback.
Crucially, the process must avoid automated judgments and retain human oversight to ensure that evaluative conclusions are made in context.
Well-designed prompts that promote accuracy, neutrality, and fairness are essential, as is a commitment to transparency about how AI is used throughout the evaluation process.
When implemented thoughtfully, AI becomes a valuable tool, accelerating the synthesis of large-scale feedback and revealing meaningful patterns, while human experts safeguard the integrity, equity, and contextual understanding of the conclusions drawn.

\section{Ensuring Proper Context in Evaluations}

A key goal of any teaching evaluation system is to ensure fairness, given the wide range of instructional settings and the well-documented biases present in student feedback.
An AI-supported evaluation system must be intentionally designed to account for instructor identity, course characteristics, and other contextual factors to avoid reinforcing unbalanced assessments.

\textbf{Instructor Identity and Bias:} Research has shown that student evaluations can reflect biases against certain instructor demographics~\cite{linse2017interpreting}.
Female instructors and those from specific racial or ethnic groups often receive lower ratings, on average, than their male or majority-group counterparts, regardless of actual teaching effectiveness~\cite{macnell2015s}.
This raises concerns when an AI system summarizes or categorizes instructors based on student feedback without adjustment, as it may unintentionally reinforce these disparities (e.g., disproportionately flagging women or minority instructors as ``needing improvement'').
To mitigate this risk, our implementation replaces each instructor's first and last name with a consistent, gender-neutral placeholder (e.g., ``Jordan Taylor'') prior to summarization.
This approach helps reduce the influence of gender or culturally specific names, which might otherwise trigger implicit biases within the model.
By standardizing the instructor's identity across all inputs and queries, the system minimizes the risk of encoding variation based on name-based assumptions.
Although this does not eliminate all forms of bias in student feedback, it removes one pathway through which identity-related bias could affect AI-generated summaries.
We also support broader bias mitigation strategies, such as including reminders that prompt students to focus on course-related feedback and remain aware of potential bias, an approach shown in research to modestly reduce biased responses. 
Although our system does not control the design of the survey, such upstream measures serve as important complements to technical safeguards.

\begin{figure}
    \centering
    \begin{tikzpicture}
[font=\small, draw=black, line width=1pt,
eval/.style={circle, draw, inner sep=0pt, minimum size=2.5mm}
]
\begin{axis}[
    width=12cm, height=7cm,
    title={Mean Scores and GPA (200-Level Courses)},
    xmin=0.3, ymax=7.5,
    ybar,
    bar width=0.9,
    ymin=0, ymax=6.5,
    xtick=data,
    ylabel={Raw Scores},
    xticklabels={Q3\\Concepts\\and Skills, Q4\\Critical\\Thinking, Q5\\Course\\Organization, Q7\\Instructor\\Feedback, Q8\\Learning\\Environment, Q9\\Pedagogy\\and Methods, Overall\\Course\\GPA},
    xticklabel style={align=center, font=\scriptsize},
    % legend title={Legend},
    legend style={at={(0.05,0.95)}, anchor=north west, font=\scriptsize, fill=white, align=left},
    grid=both, 
    major grid style={dotted,gray!40}, 
    minor grid style={dotted,gray!40}, 
    minor tick num=5
]

% % --- Quantiles and Max bands as background rectangles
\addplot [
    ybar,
    draw=darkgray,
    line width = 1pt,
    fill=white,
    bar shift=0pt
]
coordinates {(1,4.0) (2,4.0) (3,4.0) (4,6.0) (5,5.0) (6,3.0) (7,4.0)};
% \addlegendentry{Max Range}
% \addlegendimage{black, line width=1pt}
% \addlegendentry{Mean}
% \addlegendentry{Quantiles}

\addplot [
    ybar,
    draw=darkgray,
    line width = 1pt,
    fill=gray!50,
    bar shift=0pt
]
coordinates {(1,3.72) (2,3.74) (3,3.73) (4,4.68) (5,4.57) (6,2.80) (7,3.42)};

\addplot [
    ybar,
    draw=black,
    line width = 1pt,
    fill=gray!50,
    bar shift=0pt
]
coordinates {(1,3.51) (2,3.60) (3,3.35) (4,4.28) (5,4.16) (6,2.52) (7,3.19)};

\addplot [
    ybar,
    draw=darkgray,
    line width = 1pt,
    fill=white,
    bar shift=0pt
]
coordinates {(1,3.30) (2,3.35) (3,3.00) (4,3.83) (5,3.57) (6,2.21) (7,2.93)};

\draw[fill=blue!20!white, draw=black, line width=1pt] (axis cs:0.6,3.29) circle [radius=2.5pt];
\draw[fill=blue!30!white, draw=black, line width=1pt] (axis cs:0.7,3.16) circle [radius=2.5pt];
\draw[fill=blue!40!white, draw=black, line width=1pt] (axis cs:0.8,3.52) circle [radius=2.5pt];
\draw[fill=blue!50!white, draw=black, line width=1pt] (axis cs:0.9,3.44) circle [radius=2.5pt];
\draw[fill=blue!60!white, draw=black, line width=1pt] (axis cs:1.0,3.46) circle [radius=2.5pt];
\draw[fill=blue!70!white, draw=black, line width=1pt] (axis cs:1.1,3.43) circle [radius=2.5pt];
\draw[fill=blue!80!white, draw=black, line width=1pt] (axis cs:1.2,3.62) circle [radius=2.5pt];
\draw[fill=blue!90!white, draw=black, line width=1pt] (axis cs:1.3,3.42) circle [radius=2.5pt];
\draw[fill=blue!100!white, draw=black, line width=1pt] (axis cs:1.4,3.46) circle [radius=2.5pt];

\draw[fill=blue!20!white, draw=black, line width=1pt] (axis cs:1.6,2.5) circle [radius=2.5pt];
\draw[fill=blue!30!white, draw=black, line width=1pt] (axis cs:1.7,3.01) circle [radius=2.5pt];
\draw[fill=blue!40!white, draw=black, line width=1pt] (axis cs:1.8,3.15) circle [radius=2.5pt];
\draw[fill=blue!50!white, draw=black, line width=1pt] (axis cs:1.9,2.89) circle [radius=2.5pt];
\draw[fill=blue!60!white, draw=black, line width=1pt] (axis cs:2.0,3.4) circle [radius=2.5pt];
\draw[fill=blue!70!white, draw=black, line width=1pt] (axis cs:2.1,2.6) circle [radius=2.5pt];
\draw[fill=blue!80!white, draw=black, line width=1pt] (axis cs:2.2,1.7) circle [radius=2.5pt];
\draw[fill=blue!90!white, draw=black, line width=1pt] (axis cs:2.3,3.44) circle [radius=2.5pt];
\draw[fill=blue!100!white, draw=black, line width=1pt] (axis cs:2.4,2.66) circle [radius=2.5pt];

\draw[fill=blue!20!white, draw=black, line width=1pt] (axis cs:2.6,3.76) circle [radius=2.5pt];
\draw[fill=blue!30!white, draw=black, line width=1pt] (axis cs:2.7,3.73) circle [radius=2.5pt];
\draw[fill=blue!40!white, draw=black, line width=1pt] (axis cs:2.8,3.74) circle [radius=2.5pt];
\draw[fill=blue!50!white, draw=black, line width=1pt] (axis cs:2.9,3.81) circle [radius=2.5pt];
\draw[fill=blue!60!white, draw=black, line width=1pt] (axis cs:3.0,3.54) circle [radius=2.5pt];
\draw[fill=blue!70!white, draw=black, line width=1pt] (axis cs:3.1,3.75) circle [radius=2.5pt];
\draw[fill=blue!80!white, draw=black, line width=1pt] (axis cs:3.2,3.67) circle [radius=2.5pt];
\draw[fill=blue!90!white, draw=black, line width=1pt] (axis cs:3.3,3.91) circle [radius=2.5pt];
\draw[fill=blue!100!white, draw=black, line width=1pt] (axis cs:3.4,3.87) circle [radius=2.5pt];

\draw[fill=blue!20!white, draw=black, line width=1pt] (axis cs:3.6,3.89) circle [radius=2.5pt];
\draw[fill=blue!30!white, draw=black, line width=1pt] (axis cs:3.7,3.74) circle [radius=2.5pt];
\draw[fill=blue!40!white, draw=black, line width=1pt] (axis cs:3.8,3.94) circle [radius=2.5pt];
\draw[fill=blue!50!white, draw=black, line width=1pt] (axis cs:3.9,3.88) circle [radius=2.5pt];
\draw[fill=blue!60!white, draw=black, line width=1pt] (axis cs:4.0,4.24) circle [radius=2.5pt];
\draw[fill=blue!70!white, draw=black, line width=1pt] (axis cs:4.1,4.37) circle [radius=2.5pt];
\draw[fill=blue!80!white, draw=black, line width=1pt] (axis cs:4.2,4.49) circle [radius=2.5pt];
\draw[fill=blue!90!white, draw=black, line width=1pt] (axis cs:4.3,4.39) circle [radius=2.5pt];
\draw[fill=blue!100!white, draw=black, line width=1pt] (axis cs:4.4,4.58) circle [radius=2.5pt];

\draw[fill=blue!20!white, draw=black, line width=1pt] (axis cs:4.6,4.38) circle [radius=2.5pt];
\draw[fill=blue!30!white, draw=black, line width=1pt] (axis cs:4.7,4.53) circle [radius=2.5pt];
\draw[fill=blue!40!white, draw=black, line width=1pt] (axis cs:4.8,4.24) circle [radius=2.5pt];
\draw[fill=blue!50!white, draw=black, line width=1pt] (axis cs:4.9,3.84) circle [radius=2.5pt];
\draw[fill=blue!60!white, draw=black, line width=1pt] (axis cs:5.0,4.14) circle [radius=2.5pt];
\draw[fill=blue!70!white, draw=black, line width=1pt] (axis cs:5.1,3.71) circle [radius=2.5pt];
\draw[fill=blue!80!white, draw=black, line width=1pt] (axis cs:5.2,3.55) circle [radius=2.5pt];
\draw[fill=blue!90!white, draw=black, line width=1pt] (axis cs:5.3,3.68) circle [radius=2.5pt];
\draw[fill=blue!100!white, draw=black, line width=1pt] (axis cs:5.4,3.39) circle [radius=2.5pt];

\draw[fill=blue!20!white, draw=black, line width=1pt] (axis cs:5.6,2.49) circle [radius=2.5pt];
\draw[fill=blue!30!white, draw=black, line width=1pt] (axis cs:5.7,2.53) circle [radius=2.5pt];
\draw[fill=blue!40!white, draw=black, line width=1pt] (axis cs:5.8,2.34) circle [radius=2.5pt];
\draw[fill=blue!50!white, draw=black, line width=1pt] (axis cs:5.9,2.4) circle [radius=2.5pt];
\draw[fill=blue!60!white, draw=black, line width=1pt] (axis cs:6.0,2.54) circle [radius=2.5pt];
\draw[fill=blue!70!white, draw=black, line width=1pt] (axis cs:6.1,2.57) circle [radius=2.5pt];
\draw[fill=blue!80!white, draw=black, line width=1pt] (axis cs:6.2,2.43) circle [radius=2.5pt];
\draw[fill=blue!90!white, draw=black, line width=1pt] (axis cs:6.3,1.6) circle [radius=2.5pt];
\draw[fill=blue!100!white, draw=black, line width=1pt] (axis cs:6.4,1.77) circle [radius=2.5pt];

\draw[fill=red!20!white, draw=black, line width=1pt] (axis cs:6.6,2.8) circle [radius=2.5pt];
\draw[fill=red!30!white, draw=black, line width=1pt] (axis cs:6.7,2.71) circle [radius=2.5pt];
\draw[fill=red!40!white, draw=black, line width=1pt] (axis cs:6.8,2.89) circle [radius=2.5pt];
\draw[fill=red!50!white, draw=black, line width=1pt] (axis cs:6.9,3.01) circle [radius=2.5pt];
\draw[fill=red!60!white, draw=black, line width=1pt] (axis cs:7.0,2.99) circle [radius=2.5pt];
\draw[fill=red!70!white, draw=black, line width=1pt] (axis cs:7.1,3.14) circle [radius=2.5pt];
\draw[fill=red!80!white, draw=black, line width=1pt] (axis cs:7.2,2.94) circle [radius=2.5pt];
\draw[fill=red!90!white, draw=black, line width=1pt] (axis cs:7.3,3.13) circle [radius=2.5pt];
\draw[fill=red!100!white, draw=black, line width=1pt] (axis cs:7.4,3.24) circle [radius=2.5pt];

\draw[fill=white, draw=black, line width=0.5pt] (axis cs:0.55,4.75) rectangle (axis cs:2,6);
\draw[draw=darkgray, line width = 1pt] (axis cs:0.6375,5.6875) rectangle (axis cs:1,5.8125);
\draw[line width=1.5pt] (axis cs:0.6375,5.375) -- (axis cs:1,5.375);
\draw[draw=black, line width = 1pt, fill=gray!50] (axis cs:0.6375,4.9375) rectangle (axis cs:1,5.0625);

\node[anchor=west] (max) at (axis cs:1,5.75) {\scriptsize Max Range};
\node[anchor=west] (min) at (axis cs:1,5.375) {\scriptsize Mean};
\node[anchor=west] (percentiles) at (axis cs:1,5) {\scriptsize Percentiles};

\end{axis}
\end{tikzpicture}
\caption{Evaluation graphs are designed to rapidly convey feedback and highlight trends for all of the course sections taught by a single faculty member at a specific level (e.g., sophomore or 200-level).
Each bar corresponds to one question. Each circle represents a section of a course.
The departmental mean, as well as the 25th and 75th percentiles, provide context for similar courses.
For each bar, score evolution is shown from left to right, with colors progressing from lighter to darker tones.
Illustrative examples are given for typical scores (Q3), low evaluation scores (Q4), high evaluation scores (Q5), positive trend (Q6), negative trend (Q7), and a phase transition (Q8) that may reflect an adverse life event.
The average GPA (in red) is plotted alongside student feedback to provide complementary information, such as evidence of strict grading, grade inflation, or lenient grading~\cite{greenwald1997grading}.}
\label{figure:VisualReport}
\end{figure}
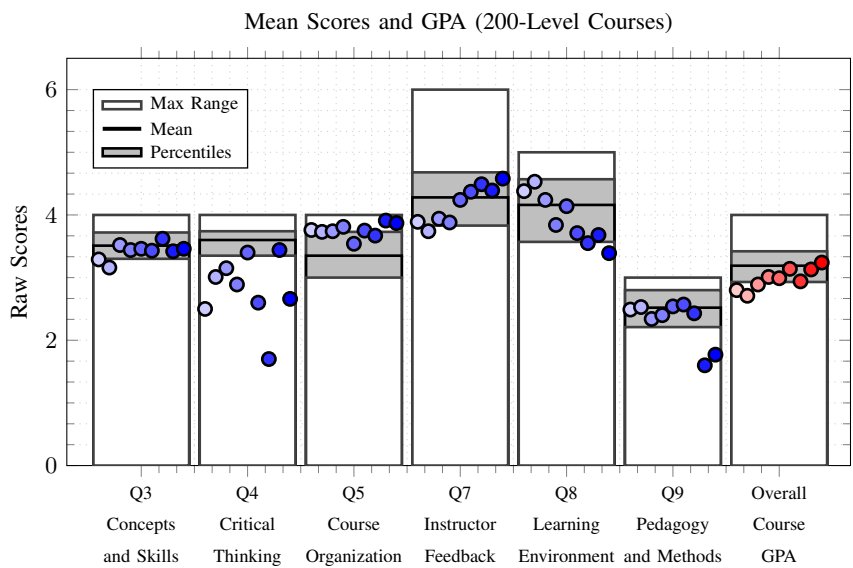

\textbf{Accounting for Class Size and Teaching Context:} The teaching context (class size, course level, required versus elective, in-person versus online, etc.) can have a significant impact on student feedback.
Large lecture courses can yield lower student satisfaction and lower evaluation scores than small, advanced offerings~\cite{johnson2013effects}.
Introductory courses or courses that students take out of requirement (like a difficult thermodynamics class for non-majors) often get more critical feedback than elective courses populated by self-selected enthusiasts. 
A fair evaluation system must recognize these differences so that instructors are not unjustly penalized for teaching challenging courses or large service courses.
Our approach uses percentile-based scoring and contextual benchmarks as one remedy.
Instead of interpreting a raw average score in isolation, an instructor's numeric scores are compared to similar courses or departmental norms in the visual report.
For example, a 4.0/5 in a class of 100 students might actually be in the top quartile for classes of that size, even if the college-wide mean is 4.3.
By providing a visual reference for large courses at a same level, the reports provide context that 4.0 is a strong result in that milieu.
Conversely, a 4.3 in a small elective might only be median when compared to other small classes that tend to score high.
This norm-referenced interpretation, illustrated in Fig.~\ref{figure:VisualReport}, helps adjust expectations and prevent one-size-fits-all judgments.
It is aligned with recommendations to include distributions and context in evaluation reports rather than just raw means. 

In addition, our visual report disaggregates the results by course type, class size, and other contextual factors.
For example, department heads reviewing student evaluations can compare an instructor's feedback in lower-division required courses separately from upper-division electives, helping to avoid ``apples-to-oranges'' comparisons.
While we do not algorithmically adjust or ``curve'' scores for large classes, we present these contextual details alongside the evaluation results to ensure they remain salient during interpretation.
Ultimately, our goal is to distinguish genuine indicators of teaching quality from contextual noise.

\textbf{Guarding Against Algorithmic Bias:} Like any AI system, large language models (LLMs) may carry latent biases if not carefully managed.
For example, models trained on internet-scale data can inadvertently associate certain genders or identities with specific descriptors.
To mitigate this risk, our system relies on AI primarily for summarization, extracting and organizing themes from student comments, rather than generating evaluative language or free-form judgments.
The summaries are grounded in the actual phrasing of the students, and the prompts are designed to promote neutral evidence-based output.
To assess potential bias, we have conducted random sampling comparisons between student comments and AI-generated summaries to ensure reasonable fidelity to the source material.
While this offers a foundational level of quality assurance, we acknowledge that further validation is needed.
In future iterations, we plan to involve instructors in reviewing and refining summaries to introduce an additional layer of human oversight.
Notably, over the past three years, we have observed improvements in summary quality as newer language models have been adopted, and we continue to evolve our system accordingly.
Although we have not identified clear instances of biased language in summaries to date, ongoing monitoring, including periodic testing with synthetic cases, remains an essential part of our governance framework.

In sum, ensuring fairness requires addressing known biases and contextual factors at multiple stages: during data collection (e.g., including bias disclaimers and carefully designing survey questions), during analysis (e.g., applying anonymization, making appropriate comparisons, and incorporating peer review data), and during interpretation (e.g., training decision-makers to consider context and providing visualizations that make that context explicit).
When thoughtfully implemented, an AI-enhanced system can help promote fairness rather than compromise it.
The objective is to ensure that every instructor receives meaningful feedback, that those who excel in challenging teaching contexts are appropriately recognized, and that no instructor is disadvantaged based on personal identity or instructional assignment.
The system's design, in tandem with clear policy guidance, is intended to uphold this principle.

\section{Visual Analytics, Feedback Loops, and Cultural Impact}

\textbf{Visualizing Teaching Effectiveness:} Humans are naturally drawn to visual information, patterns and outliers often stand out more clearly in graphics than in tables of numbers~\cite{cleveland1984graphical,ware2012information}.
For example, plotting an instructor's course evaluation scores over several years can reveal meaningful trajectories, such as whether improvements follow targeted interventions~\cite{marsh1991students}.
We include historical performance trends so that an instructor categorized as ``needs improvement'' might still recognize positive momentum, such as consistent year-over-year gains, even if scores remain below average.
This can be both informative and encouraging.
Research in educational data mining underscores that visual reports help stakeholders identify patterns that may be overlooked in raw data and make insights more accessible to those without advanced statistical training.
In fact, visualization can enable instructors to effortlessly interpret and apply teaching data by highlighting key trends in an intuitive format.
For instance, a two-dimensional plot might reveal that an instructor's scores are consistently high in smaller classes but decline in larger ones, suggesting that class size may be a contributing factor.
Recognizing patterns is a valuable first step toward instructional improvement, such as adapting pedagogical strategies for large-lecture environments.

\textbf{Percentile-Based Scoring:} As illustrated in Fig.~\ref{figure:VisualReport}, our visual reports present SET scores using quantiles and distributional context.
This approach not only improves fairness, but also supports instructor motivation.
For example, learning that one's student satisfaction scores are near the mean may encourage efforts to reach a higher quantile, such as the 75th percentile, by experimenting with new teaching strategies.
Conversely, seeing oneself in a top percentile can be validating; instructors may feel recognized as high performers (``award-worthy''), especially when raw score differences (e.g., between a 4.5 and 4.7 average) are otherwise difficult to interpret without context.
We take care to avoid using percentile rankings in a punitive or overly competitive manner.
Visual reports are shared privately with instructors and relevant evaluators, not as public rankings.
The primary goal is to support self-reflection and professional growth.
Although motivation studies suggest that clear goals and benchmarks can drive improvement, we are mindful of avoiding a gamification mindset.
The narrative surrounding these analytics is centered on learning from the data to support individual development, not competing with peers.

\begin{figure}
    \centering
    \input{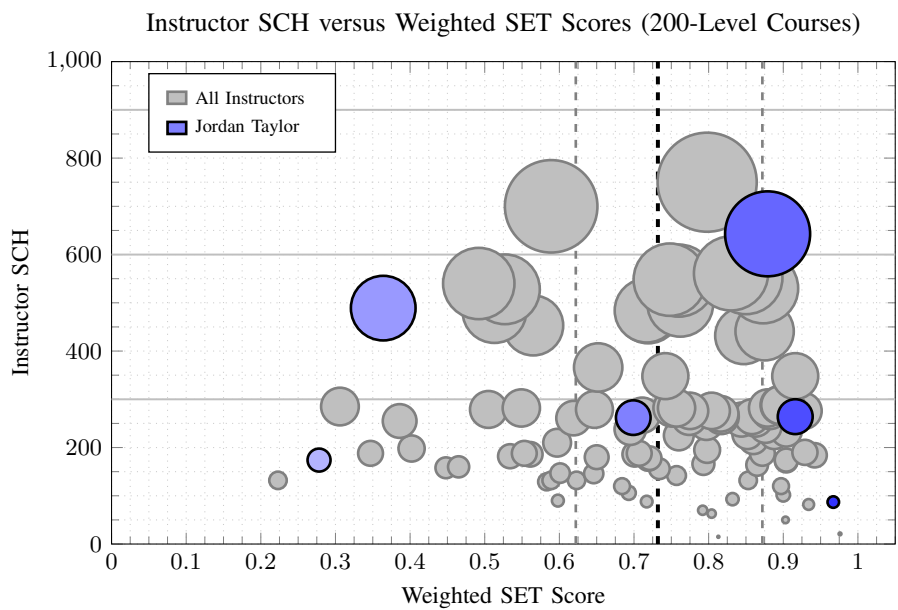}
\caption{This figure seeks to highlight teaching impact and extraordinary contributions to the academic mission within one department.
The average weighted SET score is used as a proxy for perceived course quality, whereas student credit hours (SCH) total captures instructor reach.
The size of the marker is also determined by section size and course type.
The scatter plots places average SET scores in the context of student load, with the gray circles representing the performance of other instructors within the department teaching courses at a similar level.
As before, the progression from lighter to darker tones captures evolution over time. 
}
\label{figure:ScatterPlot}
\end{figure}

\textbf{Teaching Excellence versus Impact:}
Within the College of Engineering, the evaluation of teaching intentionally incorporates two complementary dimensions: teaching excellence and teaching impact.
Excellence focuses on the quality of the pedagogical environment experienced by students and the extent to which it supports their learning.
Still, instructors teach vastly different numbers of students depending on course and section size.
Providing a meaningful educational experience in large-class settings is especially important, as it affects many more students—and, empirically, adapting or innovating instruction at scale presents unique challenges.
For this reason, it is essential to recognize the impact of teaching, which can be operationalized as strong SET scores along with a large number of students taught.
Highlighting faculty whose contributions exemplify the highest standards of instructional quality and student engagement, especially in high-enrollment or high-load contexts, ensures that both quality and reach are valued.
The design of the visual report explicitly seeks to elevate instructors whose teaching demonstrates extraordinary impact, whether through exceptionally high student credit hours (SCH), demanding course loads, or successful pedagogical strategies implemented at scale. We first summarize all of the questions in the student survey for a particular instructor in a particular section of a course with a single weighted average:
\begin{equation*}
\text{Weighted SET Score} = \frac{1}{\text{Number of Questions}} \sum_{i \in \text{Questions}}
\frac{ \text{score}_i - \text{min}_i }{ \text{max}_i - \text{min}_i } .
\end{equation*}
Then, as shown in Fig.~\ref{figure:ScatterPlot}, we plot the weighted SET score compared to semester credit hours to highlight high scores in large classes.

\textbf{Relation to Instructor Motivation:} The introduction of these tools can influence faculty attitudes in complex ways.
On the one hand, greater transparency and access to support can be highly motivating.
When instructors see their efforts acknowledged through positive summaries or upward trends and know that resources are available when challenges arise, it can boost morale and reinforce their commitment to teaching.
Although the system does not generate explicit classifications, it is designed to help department heads interpret the data and guide developmental conversations.
However, increased visibility of data can cause anxiety if not properly framed.
Some faculty might worry about constant scrutiny or become reluctant to try new teaching approaches for fear of lowering their scores.
To mitigate this, we emphasize that the system is developmental.
The culture of continuous improvement is further supported through workshops where faculty, including top performers, share effective practices and reflect on feedback in a collegial environment.
This helps shift the emphasis from evaluation to collaborative professional growth.
Research has shown that clear and constructive feedback often leads instructors to take meaningful steps to enhance their teaching. 

\textbf{Departmental and Institutional Decision-Making:} For department heads and committees, visual analytics and AI summaries can jointly help streamline decision-making processes.
Awards committees, for example, must assess the teaching effectiveness of candidates.
Rather than reviewing hundreds of raw student comments across multiple courses and years, they can consult a concise, anonymized summary that highlights consistent themes (e.g., ``Students frequently praise the instructor's well-structured lectures and concern for student learning.
A recurring suggestion is to improve feedback timeliness on assignments'').
Furthermore, committees can rapidly make sure that comments and summaries are consistent with SET scores.
This not only saves time but also reduces the undue influence of isolated or outlier comments.
It encourages committees to focus on patterns over time rather than overemphasizing feedback from a single course.
The inclusion of data on teaching load and class size further supports fairness in administrative decisions.
For example, if two candidates have comparable evaluation scores, but one taught twice as many students or more demanding courses, that context is readily visible and can be appropriately considered.
Some faculty and institutional stakeholders have expressed concern that metrics might be misused to automate personnel decisions, an outcome we explicitly aim to avoid.
Our approach emphasizes that data should inform, not replace, human judgment.
Decision-makers are strongly encouraged to interpret visual reports and AI-generated summaries in conjunction with peer evaluations, self-reflections, and other forms of evidence.

At the institutional level, aggregated analytics can inform policy and resource allocation.
For example, if an engineering college observes that large, first-year courses consistently receive lower engagement scores, this may prompt investment in instructional development for large-class pedagogy or efforts to create more manageable section sizes.
Similarly, if certain departments show improvement after adopting new teaching methods, those successes can be recognized and modeled across the institution.
Institutional research units can also publish anonymized, aggregated reports that highlight trends in teaching effectiveness metrics over time, serving as both a point of pride and a mechanism for accountability.
This contributes to a culture in which teaching quality is actively measured and valued rather than overlooked.
The literature on engineering education emphasizes that visual analytics, when implemented in academic settings, tend to elevate the visibility of what is being measured, thus signaling its institutional importance.

\textbf{Cultural Implications:} The introduction of AI-driven evaluations and visual reports conveys an important cultural message: Our college is committed to teaching excellence and educational innovation.
Engineering schools have traditionally emphasized research output, and teaching evaluations were often treated as a formality.
This initiative, by applying the same analytical rigor to teaching data as to research metrics, helps elevate the visibility and value of instruction.
Faculty see that excellent teaching can be identified and recognized (e.g., through data-informed award nominations), while areas of concern are not ignored, but met with structured support.
The objective is to foster improvement and ensure fair, evidence-based consideration for all instructors.
Shared governance played a key role in shaping this system, including the selection of evaluation questions, helping to build faculty trust and buy-in.
At this point, many department heads view end-of-term summaries as a useful synthesis of feedback, supporting more informed and constructive discussions.

By including historical data and personal trends, the system encourages a mindset of long-term growth.
Newer faculty can see their trajectory (maybe going from needs improvement to effective over a few years) which normalizes the idea that teaching is a skill honed over time, not a fixed trait.
Departments can use these data to pair mentors, e.g., a faculty with consistently high engagement scores mentoring one who is struggling in that area.
This encourages collaboration over competition.
The visual nature of the feedback also makes it easier to discuss in faculty development seminars.

Additional benefits of well-synthesized SET reports include early detection of shifting trends and sudden transitions.
The timely recognition of negative trends allows department heads and reviewers to engage with instructors before concerns escalate into major issues, enabling proactive intervention, self-reflection, and support before situations adversely affect annual evaluations.
Visual analytics also makes it possible to quickly spot atypical transitions, such as a sudden drop in SET scores for an instructor who had previously been consistent.
Such abrupt changes may indicate the impact of an adverse life event or other underlying challenges.
Recognizing these phase transitions invites more nuanced and compassionate conversations and can alert administrators to situations where a faculty member might need temporary support but may be reluctant to reach out directly.
In this way, effective SET reporting not only facilitates timely, constructive feedback, but also helps foster a culture of care and early intervention within the academic community.

\section{Validation and Benchmarking}

Implementing an AI-driven summarization and evaluation system in faculty assessment raises a critical question: \textbf{How do we know it is accurate and effective?}
Validation and benchmarking are essential to ensure that the AI-supported approach withstands scrutiny and delivers meaningful value. 
One form of validation involves comparing AI-generated summaries of student comments to those produced by expert human reviewers, such as faculty committee members or educational researchers. 
In early trials, we conducted these comparisons and found that AI summaries generally captured the same dominant themes as human readers, with substantial overlap.
These findings are consistent with those of Fuller et al.\ who reported high agreement between ChatGPT-derived themes and instructor-identified themes in course evaluations~\cite{fuller2024exploring}

Faculty acceptance is another key dimension of validation.
We solicited feedback from instructors on whether AI-generated summaries accurately reflected their course evaluations.
Most responded positively, affirming the summaries as faithful representations of student feedback. 
Some noted that the phrasing could occasionally come across as blunt.
To enhance trust and transparency, instructors and reviewers retain access to the original student comments underlying any summary, ensuring that both instructors and evaluators can verify the source content.
Notably, no faculty reported significant inaccuracies in the summaries, an encouraging indicator of the system's reliability.

We approach validation as an ongoing, multi-faceted process.
Early results suggest that AI-generated summaries align closely with human analysis, and our internal benchmarks indicate strong agreement between the system's outputs and the judgments of experienced evaluators.
Continued validation through comparative studies, faculty feedback, and long-term outcomes tracking will help ensure that the AI-supported evaluation model remains credible and reliable.
If the system fails to capture critical issues that human reviewers identify, such instances will serve as important signals to refine the process or adjust its application.
Pragmatically, success must be grounded in empirical evidence.
Our aim is not to determine whether an AI-generated summary can surpass the insight of a human expert who invests significant time reviewing student comments, drawing on principles from engineering education, and crafting tailored feedback.
Instead, we seek to elevate the annual evaluation process for every faculty member in the College, specifically in relation to teaching, and to enhance the educational experience of engineering students at scale.

In sum, visual analytics and timely feedback amplify the impact of evaluation data, making information more understandable and actionable.
They enable data-informed decisions at multiple levels, from individual teaching adjustments to college-wide policy changes, reinforcing a culture where teaching effectiveness is continually monitored and enhanced.
As one possible outcome, we envision that teaching quality in the engineering programs will steadily improve or, at least, adapt gracefully to a changing education landscape.
An important aspect of developing meaningful feedback to faculty is acknowledging the variability in student ratings for the same instructor over time.
These fluctuations highlight the need for longitudinal analysis rather than isolated judgments, and reinforce the importance of framing evaluation as part of a developmental trajectory.
Equally critical is the message sent back to students: that their feedback is not only collected but actively used to improve teaching and learning.
By demonstrating that student evaluations inform faculty development and help sustain a high-quality learning environment, we hope to reinforce the value of participation.
Over time, this may encourage greater student participation in the evaluation process.
Participation rates themselves are a measurable outcome that we intend to track, with the goal of cultivating a culture where students recognize their role in shaping educational quality.

\textbf{Survey Design and AI Summarization:}
At Texas A\&M University, the design of the Student Evaluation of Teaching (SET) survey reflects a strong commitment to shared governance and evidence-based educational assessment.
The SET instrument was developed by a university-level committee comprising educational experts and representatives from various colleges, ensuring that a range of disciplinary perspectives and teaching contexts were considered.
The resulting survey consists of twelve core questions designed to capture key aspects of teaching effectiveness and learning environments.
To promote flexibility and local relevance, the system also allows colleges, departments, and even individual instructors to add supplemental questions tailored to their unique needs and priorities.
For college-level evaluations, a dedicated committee selected six of the twelve institutional questions as most directly relevant to instructor performance, while the remaining questions address areas such as facilities and learning resources.
Student responses to these questions are collected via multiple-choice items with heavily quantized (i.e., discrete and limited) response options. The six evaluation questions used in this study are listed below.
\begin{itemize}
    \item Q3: This course helped me learn concepts or skills as stated in course objectives/outcomes.
    \item Q4: In this course, I engaged in critical thinking and/or problem solving.
    \item Q5: Please rate the organization of this course.
    \item Q7: Feedback in this course helped me learn. [...]
    \item Q8: The instructor fostered an effective learning environment.
    \item Q9: The instructor's teaching methods contributed to my learning.
\end{itemize}

\begin{figure}
    \centering
    \begin{tikzpicture}
\begin{axis}[
    width=7.5cm,
    height=7.5cm,
    matrix plot,
    colorbar,
    point meta min=-1,
    point meta max=1,
    % colormap name=viridis,
    % colormap/hot,
    colormap/jet,
    enlargelimits=false,
    axis on top,
    xtick=data,
    ytick=data,
    xticklabels={Q3, Q4, Q5, Q7, Q8, Q9},
    yticklabels={Q3, Q4, Q5, Q7, Q8, Q9},
    xtick align=center,
    ytick align=center,
    enlarge y limits={abs=0.5}, enlarge x limits={abs=0.5},
    tick label style={font=\small},
    colorbar style={tick label style={font=\footnotesize}},
    every node near coord/.append style={font=\footnotesize},
    nodes near coords,
    nodes near coords align={center},
]

\addplot+[
    matrix plot*,
    point meta=explicit,
    mesh/rows=6,
    mark=none,
] table[meta=c] {
x y c
0 5 0.687
0 4 0.713
0 3 0.829
0 2 0.749
0 1 0.636
0 0 1.000
1 5 0.389
1 4 0.413
1 3 0.547
1 2 0.409
1 1 1.000
1 0 0.636
2 5 0.599
2 4 0.677
2 3 0.663
2 2 1.000
2 1 0.409
2 0 0.749
3 5 0.621
3 4 0.654
3 3 1.000
3 2 0.663
3 1 0.547
3 0 0.829
4 5 0.936
4 4 1.000
4 3 0.654
4 2 0.677
4 1 0.413
4 0 0.713
5 5 1.000
5 4 0.936
5 3 0.621
5 2 0.599
5 1 0.389
5 0 0.687
};

\end{axis}
\end{tikzpicture}
    \caption{This figure illustrates the correlations between SET scores for the six questions used by the colleges to inform instructor evaluation.
    To ensure statistical robustness despite heavy quantization, the analysis is based on 1,032 course sections, each with 40 or more students.
    The heat map values are notably high (red), and the accompanying PCA indicates that student responses are largely driven by only one or two significant latent variables.
    }
    \label{figure:SpearmanCorrelation}
\end{figure}
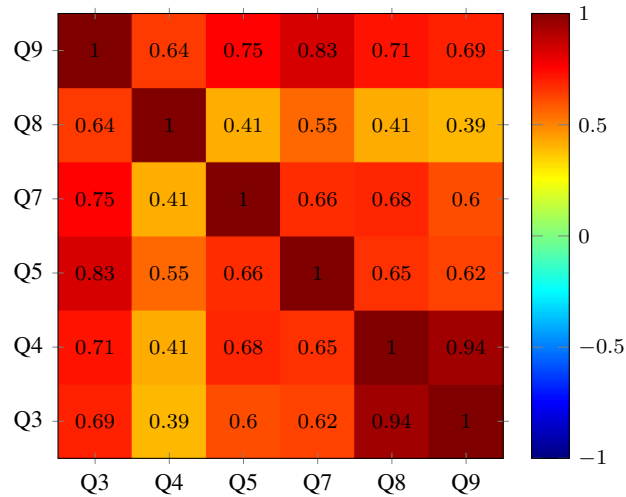

Analysis of student responses to these six questions reveals a high degree of correlation among items, suggesting substantial redundancy in the survey instrument.
Spearman correlation coefficients between normalized question averages are consistently high, with most pairs exceeding 0.6, as illustrated in Fig.~\ref{figure:SpearmanCorrelation}.
Principal component analysis (PCA) further confirms this redundancy: the first principal component alone explains more than 70\% of the total variance, while the first two principal components together account for approximately 86\% of the variance.
In practical terms, this indicates that responses to the six questions essentially reside on a two-dimensional manifold within the six-dimensional response space.
As a result, one or two latent variables are essentially sufficient to capture the meaningful variation in student sentiment, while the remaining components contribute negligible unique information.
This strong collinearity implies that the six questions are, to a large extent, measuring overlapping aspects of student experience, most likely reflecting a general construct such as perceived course quality or overall satisfaction.

In this context, and given the rapidly advancing capabilities of AI, a key question emerges: \textbf{How can SET surveys be redesigned to capture richer student feedback without increasing respondent burden?}
This analysis suggests that shorter surveys—with fewer multiple-choice items and thoughtfully constructed open-ended prompts—may yield more pertinent insights, particularly as large-scale text summarization and sentiment analysis become increasingly accessible.
However, including multiple questions can also enrich discussions within evaluation committees, fostering deeper conversations about what constitutes excellent teaching.
While this consideration falls more within the realm of human behavior than statistics or information theory, it remains highly significant.
Determining the optimal survey design is beyond the scope of this article, but it is an important and timely area for future research.
It is worth emphasizing that the AI processing pipeline described above can serve as a foundation for developing AI-aware survey instruments.

\section{Risks and Limitations of AI in Teaching Evaluation}

While the prospects of AI-aided teaching evaluation are exciting, it may be helpful to acknowledge the risks, limitations, and potential unintended consequences of integrating AI into this sensitive domain.
Identifying these issues allows us to design safeguards and governance measures to mitigate them.

\textbf{Over-Reliance and Automation Bias:} One potential risk is that committees or administrators may place undue trust in the AI-generated summaries without critically engaging with the underlying data, a phenomenon known as automation bias. 
This occurs when users assume that machine-generated output is inherently accurate and sufficient.
For example, if a committee were to rely solely on an AI summary to make decisions about a faculty member's performance without examining additional context such as course characteristics or anomalies in numerical ratings, it could lead to misguided conclusions.
To mitigate this, we provide explicit guidance and training, emphasizing that the AI summary is a starting point, not an authoritative opinion.
Our process requires evaluators to form their own independent judgments, with the AI serving as one input among several documents.
In practice, administrators receive access to comprehensive evaluation data, including distributional data and access to original student comments.
By reinforcing human oversight and encouraging active engagement with the full dataset, we minimize the risk of over-reliance on automated output~\cite{prince2004does}.

\textbf{Faculty Distrust or Reduced Agency:} Some faculty may feel alienated by the presence of an AI intermediary in the evaluation process. 
Concerns may arise that a ``black box'' is reducing the complexity of their teaching to a few lines, potentially overlooking important context or misrepresenting their efforts. 
This can lead to distrust in the process or a sense of diminished agency.
To address these concerns, we also offer training sessions to explain how the AI works, helping to demystify the technology.
Just as importantly, we emphasize that the AI summary is not definitive; for instance, during annual evaluations, faculty members are encouraged to provide contextual reflections and to engage in discussions with department chairs.
By positioning the AI system as a tool for improvement rather than judgement, our aim is to make it a resource that faculty can use to support their teaching goals.
Over time, as instructors recognize that the summaries generally reflect their own impressions from reading student feedback, we anticipate trust in the process to grow.
 
\textbf{Overemphasis on Quantifiable Aspects:} There is a philosophical concern that relying on AI and data could make the evaluation of teaching overly metrics-driven potentially overshadowing qualitative and holistic judgment. 
Some aspects of effective teaching, such as enthusiasm, mentorship, and long-term impact, may not be fully captured in student comments or quantitative scores.
If committees rely too heavily on neatly summarized data, they risk overlooking these important but less easily measured dimensions.
To mitigate this, we explicitly remind evaluators that teaching effectiveness is inherently multidimensional and that our system is designed to surface student perspectives, just one facet of a broader evaluation. 
Faculty teaching portfolios typically also include other forms of evidence, such as syllabi, teaching statements, and documented pedagogical innovations, to provide a more complete picture.
Although future system updates may allow annotating additional contributions, we emphasize that the evaluation of teaching remains a fundamentally human endeavor: informed by data but not reducible to it.
This principle is reinforced through training for all users of the system.

\textbf{Remaining Limitations:} Despite careful design and safeguards, AI-generated summaries have inherent limitations.
Student evaluations are, by nature, an imperfect proxy for teaching effectiveness; AI cannot correct for fundamental issues such as student ratings based on entertainment value or personal preferences.
If the underlying feedback lacks depth, the resulting summary will inevitably reflect that superficiality.
Additionally, there can be a disconnect between student perceptions and actual learning; innovative instructors who challenge students intellectually may receive average evaluations, even while delivering outstanding instruction.
Such contributions may only be fully recognized through longer-term outcomes or external validation.
For these reasons, we are careful not to rely exclusively on AI-generated outputs for any high-stakes decisions.
Rather, they are treated as one piece of a broader, multi-dimensional evaluation process.

The use of AI in teaching evaluation introduces risks, but each can be addressed through thoughtful design and sound policy.
By proactively identifying potential challenges (such as bias or over-reliance) and implementing layered safeguards, we strive to leverage the strengths of AI (scale, speed, consistency) while minimizing unintended consequences.
Ongoing monitoring and a readiness to adapt are essential.
Our guiding principle is to ensure that AI supports, rather than undermines, our educational mission and core values.
Thus far, this cautious approach has been effective: we have not encountered significant issues and remain vigilant as the system continues to scale.
Maintaining this reflective, responsive posture will be critical as AI tools grow in capability and influence.

\section{Implications for Recognition and Faculty Development}

The deployment of an AI-supported teaching evaluation system at scale carries broader implications for institutional policy, faculty recognition, and professional development infrastructure.
In many ways, it challenges us to rethink and refine how we value and support teaching in an engineering-focused institution.

\textbf{Institutional Policy and Evaluation Standards:} An important next step is to update institutional policies on faculty evaluation to thoughtfully incorporate AI-generated insights while ensuring their appropriate use. 
Efforts are underway to clarify how summarized student feedback produced by AI tools may be used in annual review processes.
For instance, one possibility is to discuss these considerations in the COE Faculty Evaluation guidelines that affirms the value of AI summaries as a supplement to more comprehensive evaluations, while emphasizing the necessity of human interpretation and disclosure.
Similarly, we are working to establish safeguards that prevent adverse personnel actions, such as placing a faculty member on a performance improvement plan, from being based solely on AI-generated results, requiring instead a collaborative review process involving teaching support staff and contextual evidence.
At the same time, we are exploring how exceptionally strong, consistently positive teaching summaries might be appropriately factored into merit and recognition frameworks.
Codifying these practices remains an ongoing institutional priority aimed at fostering consistency across departments, supporting fairness in the use of AI-derived data, and signaling a broader commitment to innovation in assessment.
These evolving guidelines may also serve as a reference point in accreditation and quality assurance efforts.

\textbf{Recognition and Reward Systems:} As clearer evidence of teaching excellence emerges from the data, we have strengthened our recognition programs.
For example, the selection process for teaching excellence awards in the College of Engineering now leverages analytics, highlighting faculty in the top percentiles with qualitative indicators of impact, to help identify deserving nominees.
Importantly, AI is not used to determine award recipients on its own; rather, it serves as a tool to surface potential nominees, who are then reviewed by a committee through more traditional means, such as classroom observations or teaching portfolio evaluations.
This process balances data-driven identification with qualitative judgment.
As a result, faculty who have been quietly excelling in their teaching are more likely to be recognized and celebrated, thereby providing a boost to morale and encouraging a broader culture of instructional excellence.

Summarized peer evaluation data can also support the formation of peer mentoring networks.
For instance, if evaluations suggest that one instructor could strengthen their assessment design, and another is recognized by both peers and students for excellence in that area, the two can be intentionally paired.
The Institute for Engineering Education and Innovation (IEEI) plans to use these insights to identify common challenges across the faculty (e.g., frequent concerns about providing effective feedback on assignments) and organize targeted workshops or training accordingly.
In this way, the data-driven approach helps prioritize and justify investments in teaching development, enabling a more strategic and needs-based allocation of resources.

The integration of AI in teaching evaluation at an engineering institution is more than a technical upgrade; it is a catalyst for policy evolution and cultural change~\cite{borrego2010diffusion}.
It reinforces that effective teaching is measurable, improvable, and worth rewarding.
However, it also demands robust support systems.
As we refine our approach, we foresee a future where faculty evaluation is a holistic, data-informed, and constructive process that drives both individual growth and education excellence at scale.

\section{Implementation Notes}

The implementation of our AI-supported summarization system involves a blend of structured data handling, natural language processing, and privacy-conscious design.
We integrate data from two primary sources: grade records accessed through structured SQL queries and student evaluation comments obtained from CSV exports provided by our institution's Office of Institutional Effectiveness \& Evaluation.
The incoming data often requires significant preprocessing to ensure accuracy, consistency, and relevance before it is suitable for summarization.

To prepare the data, Python-based scripts handle a comprehensive set of conditioning tasks aimed at ensuring textual consistency, privacy, and fairness in downstream summarization. 
These tasks begin with normalization routines to address encoding artifacts, such as reversing inherited \emph{mojibake} resulting from inconsistent character encodings, and proceed to standardize typographic elements like quotation marks, apostrophes, and line breaks.
Extraneous whitespace and formatting anomalies are also cleaned to ensure reliable, tokenizable inputs for summarization.
 
A critical next step is anonymization.
Since student comments often contain instructor names, sometimes spelled incorrectly or in varying formats, our system uses approximate string-matching techniques to identify and redact these references.
Specifically, we employ the \texttt{fuzz} library, which applies variants of the Levenshtein distance algorithm to measure partial string similarity between instructor names and comment text.
This allows the system to detect and replace instances of an instructor's name even when there are minor typographical errors, nicknames, or inconsistent casing.
Identified names are substituted with generic placeholders (e.g., ``Jordan Taylor''), ensuring the comments retain their pedagogical meaning while removing personally identifiable markers.
This approach protects instructor privacy and helps reduce the possibility of bias creeping into the summarization process, particularly from repeated name mentions that might unintentionally color the AI's output. 
Comments are then grouped and aggregated by instructor and section using unique identifiers, allowing the system to process coherent bundles of feedback associated with individual teaching assignments.

Once preprocessed, anonymized feedback is passed to the Gemini 1.5 Flash model (\texttt{google.generativeai} library) through a structured prompting system that guides the AI to produce summaries in a neutral, concise, and paragraph-style format.
Safety configurations are applied to block outputs that might contain hate speech or harassment, and exceptions (e.g., flagged or malformed responses) are handled programmatically. 
Each summary is logged and stored for inclusion in a faculty member's evaluation report.

For the 2023--2024 academic year, the system processed a total of 141,512 student comments across 10,733 course sections, some of which included subsections such as lab or stacked components.
This resulted in 5,375 distinct API calls to the Gemini summarization model.
For the 2024--2025 academic year, the system processed a total of 124,390 student comments across 10,995 course sections, which generated 5,191 distinct API calls to the Gemini summarization model.
As a side note, small sections with low response rates, where summaries may not be representative, are omitted~\cite{giesey2004motivation,kember2015motivating}.
The process is designed to scale and can accommodate the demands of a large engineering college without manual bottlenecks.

Throughout, the goal is to streamline the evaluation workflow while maintaining transparency, fairness, and interpretability.
The technical pipeline ensures that the AI operates on clean, anonymized input and returns summaries grounded in the language of students themselves.
While the underlying code and API logic remain invisible to most end users, the outputs are structured in a way that is intelligible and usable for instructors and evaluators alike.

\subsection*{Code Availability}

The complete set of Python scripts used to implement the AI pipeline and generate visual reports is available in our GitHub repository: \href{https://github.com/EduAdminReps/tamu-coe-setreports}{tamu-coe-setreports}.
Please note that, in accordance with institutional privacy requirements, survey data and student responses are not included in the repository.
The software is provided to enable researchers and practitioners to apply these methods to their own evaluation data.

\section{Conclusion}

The challenge of evaluating teaching effectiveness at scale, particularly in a large engineering school, has traditionally been marked by information overload, inconsistent use of data, and concerns about fairness.
Our exploration into an AI-supported evaluation system demonstrates that, when carefully implemented, such technology can significantly enhance the process. 
It synthesizes vast qualitative feedback into coherent insights, surfaces specific strengths and areas for improvement, and does so in a timely manner that can feed into continuous improvement cycles~\cite{chong2021reconsidering}.
Equally important, it provides a platform for fairness by contextualizing results and prompting multi-faceted evaluation (student, peer, self-reflection), aligning with scholarly recommendations for holistic assessment.

AI-driven summarization techniques, like hierarchical LLM summarization, have shown they can adhere to qualitative best practices, condensing without (much) distortion, and doing so under ethical guardrails like anonymization and human oversight.
In our case, the AI acts as augmenting intelligence for faculty and committees, handling the heavy lifting of data synthesis while humans bring in the nuanced judgment.
This helps maintain the integrity of the evaluation process.
It also exemplifies transparency; we have adopted guidelines to not use AI to make the final evaluative call, thereby preserving the role of human discernment in a critical human endeavor: educating students.

By accounting for course context, our system strives to promote educational excellence broadly.
It recognizes that effective teaching can look different across contexts and that evaluation must be adjusted accordingly to be fair.
Furthermore, our approach consciously monitors and mitigates biases, both those present in student feedback and those potentially introduced by algorithms.
The incorporation of visual analytics and rapid feedback mechanisms has begun to transform the culture within our institution.
Teaching effectiveness is no longer a hidden indicator; it is visible, trackable, and actionable.
Departments, armed with clearer data, can allocate support or recognition more judiciously, reinforcing a culture of continuous improvement. 

Early validation efforts, both internal and external, indicate that AI-supported systems can perform comparably to traditional evaluation models in identifying key feedback themes and teacher performance levels.
It does so with greater efficiency and consistency, though we remain cautious to keep it as a support tool rather than an oracle.
The broader implications of our approach suggest a positive trajectory: one where teaching excellence is more systematically fostered and rewarded.
In the long run, this benefits not just faculty careers, but, most critically, student learning outcomes and the overall educational mission of the university.
By identifying patterns across thousands of student voices, we can implement changes that improve curriculum and pedagogy at scale, essentially closing the feedback loop between students and the institution in a much tighter fashion than before.
Such a workflow may well become standard in large universities.
Finally, our work contributes to the evolving narrative of what it means to be a faculty member in the twenty-first century.
It underscores that teaching is a dynamic, evidence-informed practice --- one that can be continually improved with the right feedback and support~\cite{henri2017review}.

\section{Acknowledgment}

The authors are grateful to C. Cantrell, Engineering Data Service, for guidance on data aggregation.
They also acknowledge E. Piwonka and the Office of Institutional Effectiveness \& Evaluation (OIEE) at Texas A\&M University for their assistance in navigating student scores and comments.

\nocite{johri2020artificial}

\bibliography{references}
\bibliographystyle{plain}

\end{document}